# The Molecular Environment of the Gamma-ray Source TeV J2032+4130

Yousaf M. Butt[1], Nicola Schneider[2], T. M. Dame[1], Christopher Brunt[3]

[1] Harvard-Smithsonian Center for Astrophysics, Cambridge, USA

[2] DAPNIA/SAp CEA/DSM, Laboratoire AIM CNRS, Gif-sur-Yvette, France

[3] School of Physics, University of Exeter, Exeter, EX4QL, UK

**Abstract**

The mysterious very high energy gamma-ray source, TeV J2032+4130, is coincident with the powerful Cygnus OB2 stellar association, though a physical association between the two remains uncertain. It is possible that the detected very high energy photons are produced via an overdensity of locally accelerated cosmic rays impinging on molecular clouds in the source region. In order to test this hypothesis, we used the Kitt Peak 12m, the Heinrich-Hertz Submillimeter Telescope (HH-SMT), and the Five College Radio Astronomy Observatory (FCRAO), to obtain observations in the $J=1 \rightarrow 0$ and $J=2 \rightarrow 1$ lines of both $^{12}CO$ and $^{13}CO$. We report here on the detection of significant molecular material toward the TeV source region which could be acting as the target of locally accelerated CRs. We also find evidence of compact molecular clumps, showing large line widths in the CO spectra, possibly indicative of energetic processes in this region of Cygnus OB2.

I. **Introduction**

Some years ago, the High Energy Gamma Ray Astronomy (HEGRA) array of imaging Cherenkov telescopes detected an unexpected gamma-ray source, TeV J2032+4130, within the angular extent of the massive Cygnus OB2 association (Aharonian et al., 2002, 2005). This is one of the richest OB clusters in the galaxy with an estimated population of ~2600 OB stars (Knodlseder 2000) within a diameter of about 60 pc (equivalent to ~2° at a distance of 1.7 kpc). TeV J2032+4130 is spatially extended and steady in intensity indicating it is likely of Galactic origin, yet still has no firmly identified counterparts at lower energies. Although follow-up observations using CHANDRA and the VLA were inconclusive (Butt et al. 2003, 2006, 2008; Mukherjee et al., 2003, 2007; Paredes et al., 2007; Marti et al. 2007), the low X-ray and radio emission indicates that the TeV flux is likely hadronic, as opposed to leptonic, in origin[1]. Recently, the HESS collaboration detected TeV gamma-ray emission from the Westerlund 2 stellar association in the southern sky (Aharonian et al. 2007) which may also indicate turbulent particle acceleration induced by stellar shocks, perhaps with a contribution from "hidden" supernova remnants expanding in low density media (e.g. Bykov 2001; Tang & Wang 2005). If it can be shown that the TeV source is due to Cygnus OB2 accelerating cosmic rays to TeV energies, this would constitute an important step towards confirming both a new type of Galactic cosmic ray accelerator, and a new source class of gamma-ray emitter.

In this Letter, we describe high resolution CO observations in the $J=1\rightarrow0$ and $J=2\rightarrow1$ lines of both $^{12}$CO and $^{13}$CO performed in the immediate environment (±30') of TeV J2032+4130. Investigating the possible relationship between molecular gas and high-energy emission can be quite fruitful, as a recent study of the Westerlund 2 TeV source illustrates (Dame 2007). The situation for the Cygnus source is much more complicated owing to a great deal of confusion along this line of sight [see, e.g., Figure 10 of Molnar et al. (1995)]. However, due to recent CO studies by Schneider et al (2006, 2007), it is now possible to disentangle the various molecular line emission features seen in the Cygnus region and, thus, more robustly interpret the high-resolution CO data obtained in this investigation.

---

[1] A recent analysis of XMM data reports a possible region of diffuse X-ray emission roughly coincident with the TeV source (Horns et al., 2007). However, diffuse emission of the morphology reported by Horns et al. (2007) has not been detected in a deep 50 ksec Chandra exposure (Butt et al., 2006), and thus remains unconfirmed. It is likely that the possible diffuse emission reported by Horns et al. (2007) could instead due to the many X-ray point sources known to exist in the field (Butt et al., 2006) which are simply not resolvable by XMM.

## II. Observations

We observed the central 22´×22´ field around TeV J2032+4130 in the $^{12}CO(1\rightarrow0)$ line using the Arizona 12m telescope in February and March of 2007, and a region of 10´×10´ in the $^{12}CO(2\rightarrow1)$ and $^{13}CO(2\rightarrow1)$ lines using the HH-SMT 10m Telescope in May of 2007. The FCRAO $^{13}CO(1\rightarrow0)$ data (Simon et al., in prep.; Schneider et al. 2007) and $^{12}CO(1\rightarrow0)$ data (Brunt et al., in prep) were also used to understand the emission features in an extended region (1°×1°) centered on the TeV source. The 12m and SMT data are beamwidth sampled (1' and 30", respectively) while the FCRAO data have a fully-sampled gridding of 22.5".

For all molecular line data we refer to RA(2000)=$20^h32^m00^s$, Dec(2000)=41°31'00" as the (0,0) position. The various observational parameters are summarized in Table 1.

| Species | Line | Freq. $\nu$ GHz | $B_{eff}$ | HPBW arcsec | $\Delta v$ km s$^{-1}$ | rms K | Observing Period | Telescope Used |
|---|---|---|---|---|---|---|---|---|
| $^{13}CO$ | $1\rightarrow0$ | 110.201 | 0.48 | 46 | 0.20 | 0.24 | 2003-2006 | FCRAO 14m |
| $^{12}CO$ | $1\rightarrow0$ | 115.271 | 0.48 | 46 | 0.25 | 0.80 | 2006 | FCRAO 14m |
| $^{12}CO$ | $1\rightarrow0$ | 115.271 | 0.70 | 54 | 0.65 | 0.16 | 2-3/2007 | Arizona 12m |
| $^{12}CO$ | $2\rightarrow1$ | 230.538 | 0.77 | 33 | 0.15 | 0.11 | 5/2007 | Arizona SMT 10m |
| $^{13}CO$ | $2\rightarrow1$ | 220.399 | 0.77 | 35 | 0.34 | 0.11 | 5/2007 | Arizona SMT 10m |

Table 1: The observational parameters for the various data reported in this study. The columns give the species, the line transition, frequency, main beam efficiency $B_{eff}$, half power beam width (HPBW), velocity channel width $\Delta v$, rms on a $T_{mb}$ scale, the observing period, and telescope used.

## III. Results

*(a) Large-scale view: FCRAO $^{13}CO(1\rightarrow0)$ data*

The velocity integrated FCRAO $^{13}CO(1\rightarrow0)$ data in a 1 sq degree region around TeV J2032+4130 are shown in Figure 1a. In the central region around the TeV source (±10'), a cavity is observed[2], while in the outskirts of the map (typically about 20´-30´ from the center) some compact clumps are seen, embedded in lower column-density material. Three clumps, for which the associated FCRAO $^{12}$CO(1→0) spectra are shown in Fig. 1b, have been circled. The clumps appear at all velocities between -28 and 26 km/s while the bulk of the Giant Molecular Cloud (GMC) emission in Cygnus X is found between -10 and +10 km/s. It appears plausible, however, that the high-velocity clumps (i.e. those at <-10 km/s and >10 km/s) are still part of the Cygnus cloud complex but have been ejected from the region due to energetic processes.

The $^{12}$CO(1→0) spectra from the three clumps highlighted in Fig. 1a clearly show non-Gaussian line profiles with large line widths (Fig. 1b) and clump 1 even reveals a broad red wing. There appears to be no *embedded* mid-IR source in clump 1 and the emission observed in this direction with the Midcourse Space Experiment (MSX) is due to a foreground O-star [Massey & Thompson 1991]. The two other clumps show no (mid)-IR emission. Since recent SiO observations (Schneider et al., in prep.) show no outflow emission due to a young stellar object embedded in these clumps, we tentatively propose that the line wings and the presence of high-velocity, compact clumps could be due to the mechanical power injected by the massive stars of Cyg OB 2 and/or the possible expanding shock waves from one or more "dark" SNRs (e.g. Tang & Wang 2005), or perhaps even both. A more detailed study of the physical properties of the compact, high-velocity clumps will be given in a forthcoming paper (Schneider et al., in prep.).

*(b) Arizona 12m and SMT data in the TeV J2032+4130 region*
While a total column density map such as Figure 1 shows only a cavity around TeV J2032+4130, closer examination of the channel maps such as those in Figure 2 reveal emission also within the TeV source region. The weak diffuse $^{12}$CO(1→0) emission within the angular extent of the TeV source region in the velocity interval between v=6 to 14 km s$^{-1}$ in Fig. 2 is associated with the foreground (~600pc) Great Cygnus Rift (Schneider et al. 2006, 2007).

---

[2] This $^{13}$CO cavity is reminiscent of a smaller IR void found in IRAS 60$\mu$m amd 100$\mu$m data which is interestingly coincident with the TeV source. See section 3.2 of Butt et al. (2003) for further details.

An interesting extended feature coincident with the TeV source lies at velocities between v=-10 to -3 km s$^{-1}$. This gas cloud (called "TeV clump" henceforth for brevity) appears to be nestled between two regions of weak radio emission seen in our mosaiced 6cm VLA image (Butt et al., 2008), as shown in the overlay map Fig. 3. However, this relative positioning of the TeV clump and radio emission could well be a chance coincidence. This clump is also prominent in both the higher excited $^{12}$CO(2→1) and $^{13}$CO(2→1) lines (Fig. 4). This gas clump does not show prominent wings – only a weak red wing is detected but this remains to be confirmed with higher angular resolution CO line observations that are now underway. The $^{12}$CO 2-1/1-0 line intensity ratio is found to be around ~1 for this clump. This would indicate rather unperturbed molecular material and we, thus, cannot yet rule out the possibility that this clump is actually simply a fore- or background object not physically associated with the TeV source.

**Discussion**

The mass of the TeV clump as deduced by the $^{13}$CO(1→0) emission between v=-10 to -3 km s$^{-1}$ is ~ 337 M$_\odot$. This is a factor of ~5 larger than the value of ~66 M$_\odot$ derived for the mass possibly associated with the TeV source, using much coarser $^{12}$CO(1→0) and HI data in Butt et al. (2003). (That value was based on assuming that the mean gas density derived over a much larger volume around the TeV source also applied within the 6' radius of TeV source.) Assuming that the TeV source and the molecular clump are indeed related, the higher deduced mass would correspondingly reduce the local CR luminosity needed to explain the TeV source as arising from hadronic interactions. In Butt et al. (2003) a value of ~10$^{36}$ ergs/sec was derived based on the 66M$_\odot$ value; the higher mass found here, if physically related, would reduce the required luminosity to, ~10$^{35}$ ergs/sec. However, note that although a CR acceleration efficiency of only ~0.1% was proposed in Butt et al. 2003, (yielding the requisite ~10$^{36}$ ergs/sec from the value of the total Cyg OB2 power of ~10$^{39}$ ergs/sec proposed by Lozinskaya et al., 2002), this efficiency value did not take into account the small ~0.5 str solid angle subtended by the ~6arcmin radius TeV source located about 15 arcmin from the core of Cyg OB2. Taking into account the new lower value of the required CR luminosity needed at TeV J2032+4130, together with an approximate correction for the geometry yields a required CR acceleration efficiency at the core of Cyg OB2 of $\sim (5)^{-1} \times (0.5\,\text{str}/4\pi\,\text{str})^{-1} \times 0.1\% \sim 0.6\%$, which is still very reasonable. If the locally accelerated CRs are also interacting with the material near the stars in the region, as proposed by Domingo-Santamaria & Torres (2006), then the above analysis will, of course, have to be adjusted to reflect the additional target mass available, further reducing the required CR acceleration efficiency of Cyg OB2.

If the putative CRs accelerated by Cyg OB2 are, however, more uniformly accelerated throughout the association, and do not emanate just from the central region, then the solid-angle argument must be modified to take into account the volume ratio of the TeV source region to that of entire association, which would yield a required efficiency of 6% instead of the 0.6% value derived above. This larger value would still be consistent (and even somewhat less) than the predicted values of shock-driven CR acceleration efficiencies of SNR shockwaves, which typically range ~10-30%.

Why would it be that only the TeV clump appears TeV bright and that the other clumps within Cyg OB2 do not? The easiest explanation may be that although the central velocities of all the clumps in Fig. 1 are *consistent* with the distance to Cyg OB2, only the TeV-clump may, in fact, be at the precise distance of Cyg OB2. Another possibility is that, as in the scenario outlined in Section 4 of Butt et al. (2006), any intrinsic TeV emission produced in the more centrally located molecular clumps may be being attenuated via pair-production in the optically and IR-brighter central regions of Cyg OB2 (e.g. Moskalenko et al., 2006).

If the TeV emission is produced via pion decay then one can estimate the radio synchrotron emission from the secondary leptons, and compare this with the rough value measured by the moasiced VLA observations of the region at 6cm (Butt et al., 2008). As argued in Section 4 of Butt et al. (2003), it is likely that the secondary electrons have reached steady state. We may simply rescale the radio curves in Fig. 10 of Butt et al. (2003) – derived for an assumed magnetic field of ~5$\mu$G – to the newly-derived density of the TeV source region by using the parametric scaling: $\sim (B/5\mu G)^{1.5}/(n_{tot}/30\text{cm}^{-3})$. As mentioned above, the newly-derived density of the TeV source region (of course, under the assumption that the TeV-clump is indeed the source of the TeV gammas) is a factor ~5 greater than that assumed in calculating Fig 10 in Butt et al. (2003) and thus the predicted synchrotron emission of the secondaries at 6cm (4.9 GHz) is ~200mJy/5=40 mJy, for the same magnetic field of 5 $\mu$G used in those simulations. However, the measured diffuse 6cm flux in this region is, instead, approximately 400mJy (Butt et al., 2008), roughly ten times larger. Thus, one possibility is that the magnetic field of the TeV clump may be ~25$\mu$G, since this would yield the requisite factor of 10 ~$(25\mu G/5\mu G)^{1.5}$ increase in the predicted radio synchrotron intensity. Indeed, this 25$\mu$G value would be much more in line with measured magnetic field values of Galactic molecular clouds, typically ~10-100$\mu$G (e.g. Stahler & Palla, 2004).

Lastly, we emphasize that we cannot yet confirm that the TeV-clump is physically related to the TeV emission and it may well be merely a chance coincidence; we strongly encourage deeper multiwavelength observations to either confirm or discount this possibility.

**Acknowledgement**

We are very grateful to the director, Lucy Ziurys, and personnel at the Arizona Radio Observatory (operated by University of Arizona): Tom Folkers, Bob Moulton, Mike Begam, Erin Hails and Teresa Longazo. We thank R. Simon for providing $^{13}$CO(1→0) FCRAO data. YB is partially supported by a NASA Long Term Space Astrophysics grant.

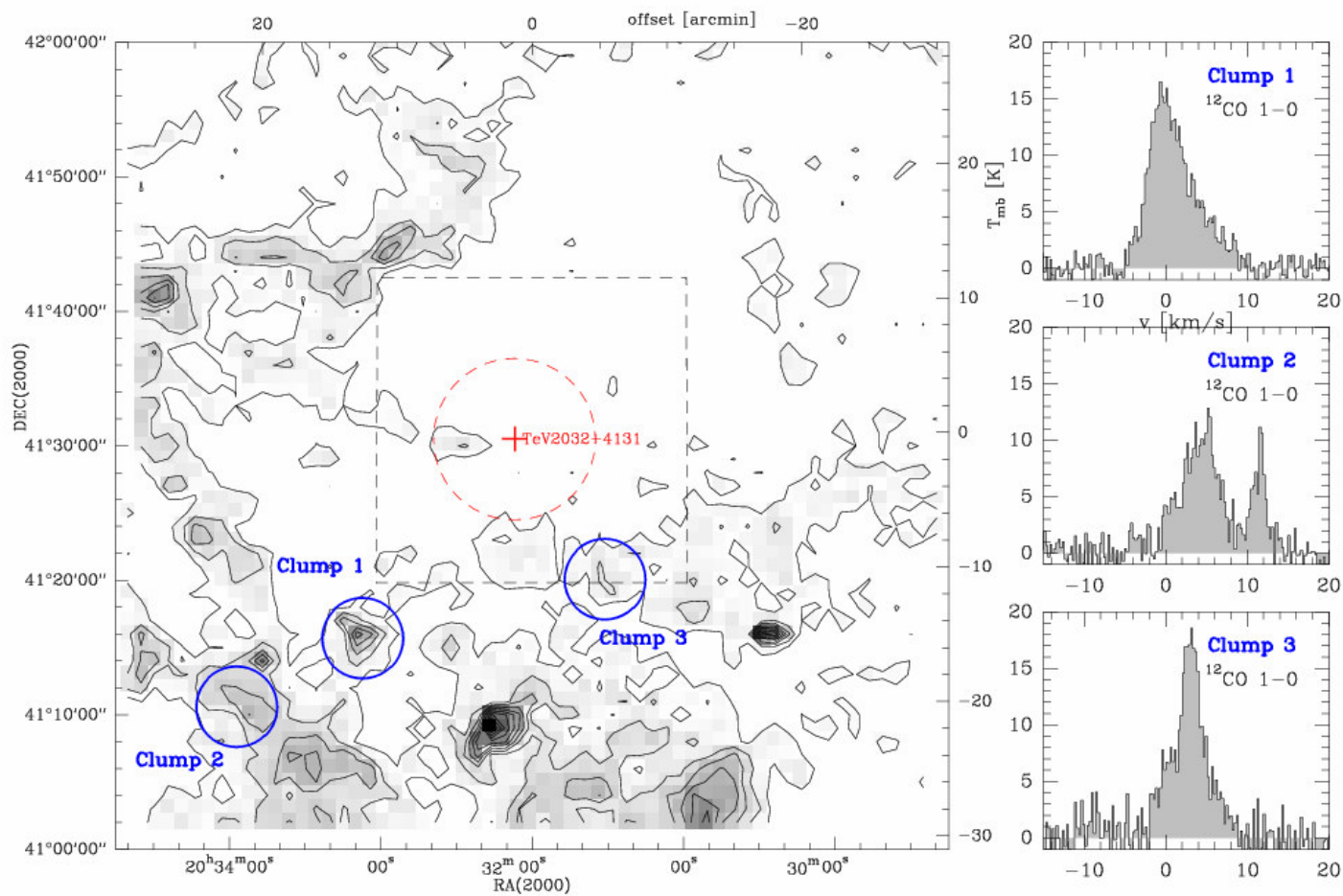

**Figure 1a (left):** FCRAO $^{13}$CO (1→0) emission in the velocity range -28 to 26 km/s around the gamma-ray source TeV J2032+4130 (the center of gravity position is indicated by a red cross and the Gaussian fitted extent of ~12´ indicated by a red circle). The contour levels range from 1.325 K km/s to 25.175 K km/s in steps of 2.65 K km/s. Some gas clumps detected in the field are marked with blue circles. [Though these clumps are not all very prominent in the integrated velocity map shown here, they have been confirmed to be discrete compact clumps in channel maps which are not shown here.] **1b right**: FCRAO $^{12}$CO (1→0) spectra from the center positions of the clumps circled in Fig.1a.

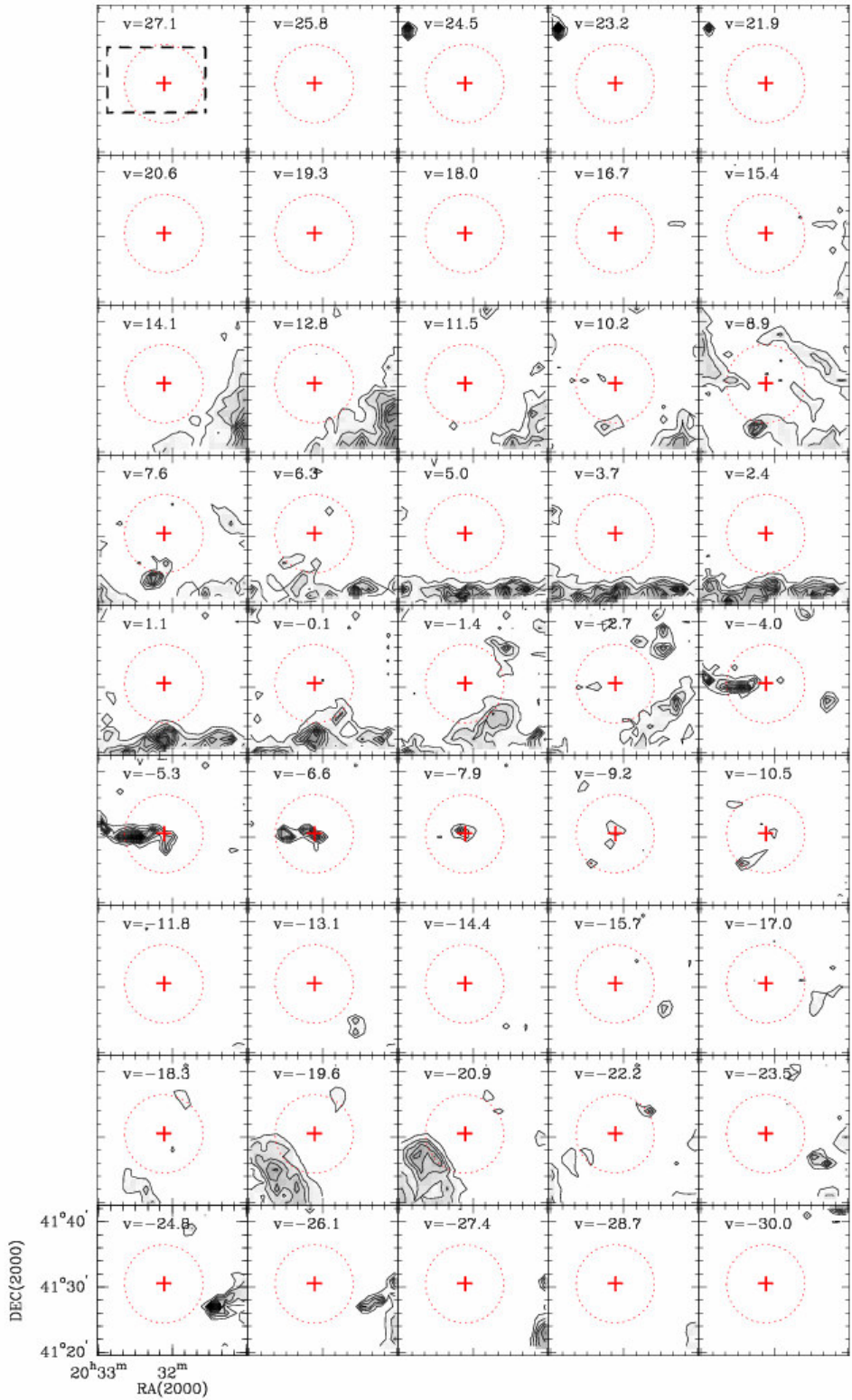

**Fig 2:** Channel maps of $^{12}$CO (1→0) emission, obtained with Kitt Peak, in the velocity range ~27 to ~ -31 km/s around the gamma-ray source TeV J2032+4130 (indicated by a red circle). The contour levels range from 0.6 K km/s to 10.2 K km/s in steps of 0.8 K km/s. In the first panel, the region observed in the $^{12}$CO and $^{13}$CO (2→1) lines with the SMT is outlined by a dashed square.

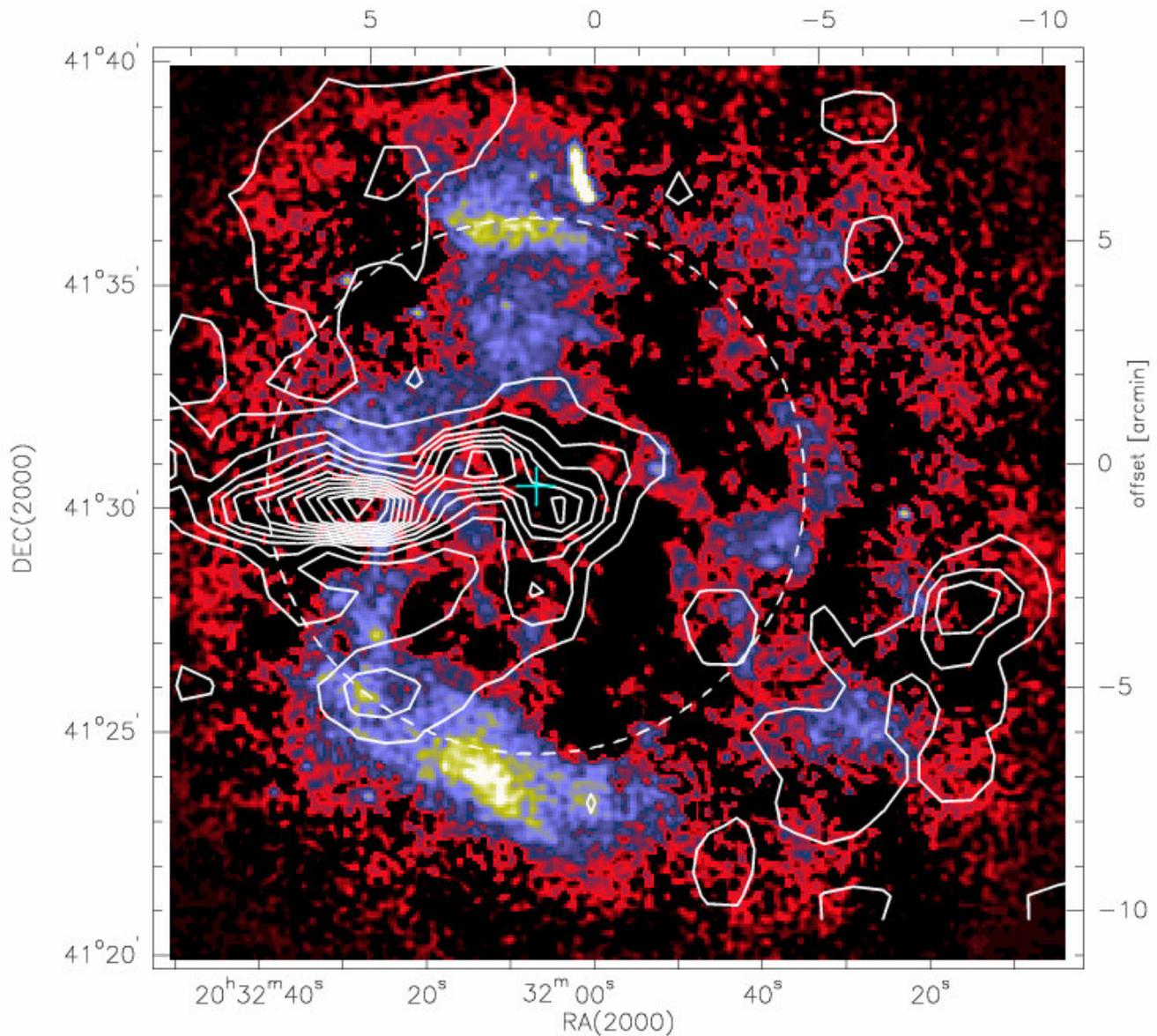

**Fig 3:** Contours of Kitt Peak $^{12}$CO (1→0) emission in the velocity range -15 to -4 km/s overlaid on a mosaicked color-scale image of 6 cm emission obtained with the VLA (Butt et al., 2008). The contour levels range from 1.32 K km/s to 25.52 K km/s in steps of 2.2 K km/s. The ring-like radio structure detected at 6cm, coincident with the TeV source, is partially non-thermal and may, in fact, be a plausible counterpart of the TeV source (Butt et al., 2008; Marti et al., 2007).

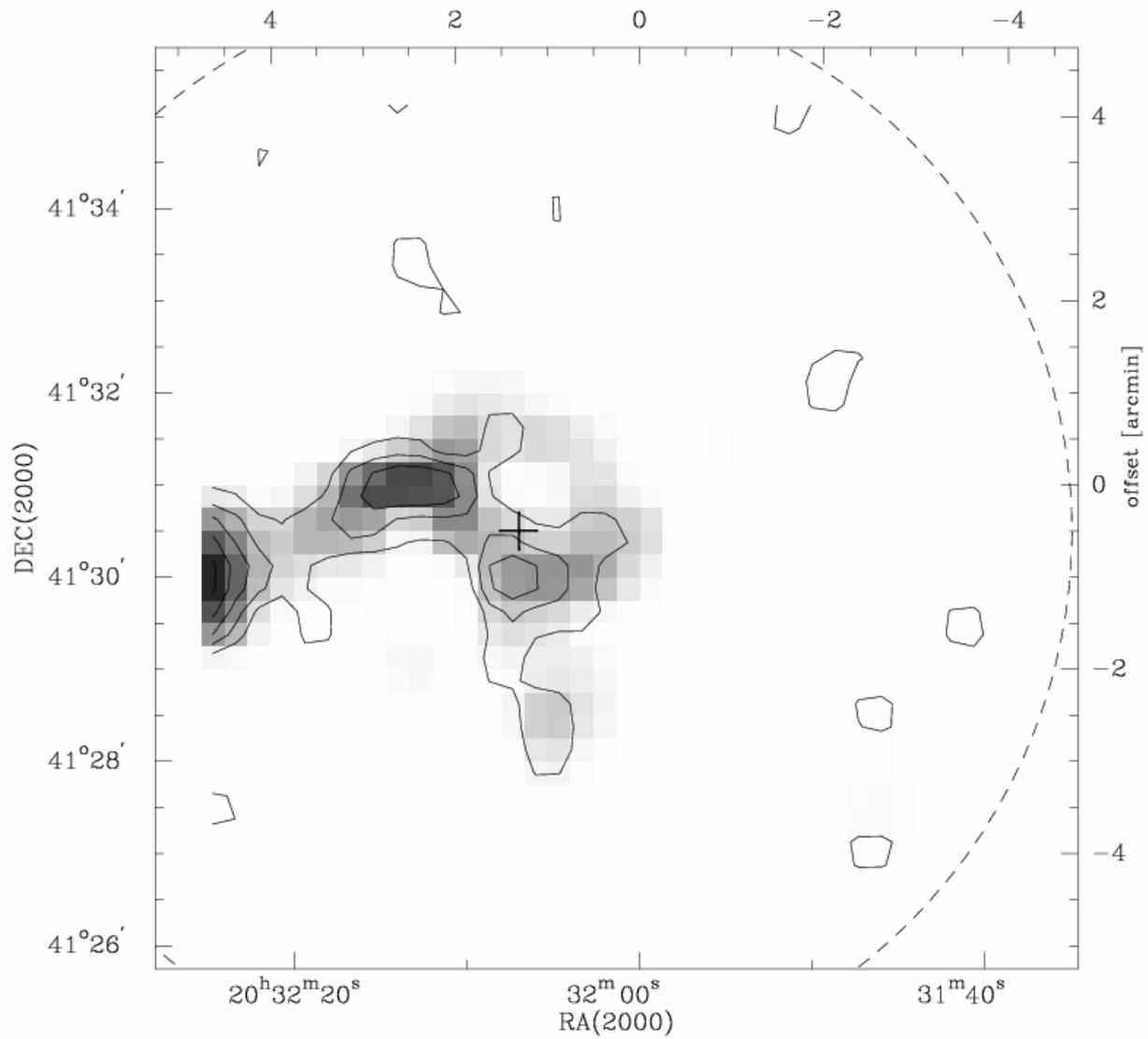

**Fig 4:** The molecular environment around TeV J2032+4130 in the velocity range -15 to -4 km/s in the $^{12}$CO (2→1) line (gray scale from 0.1 to 0.5 K km/s) and $^{13}$CO (2→1) line (contours from 0.05 to 0.25 K km/s in steps of 0.05 K km/s).